\documentclass{aa}
\usepackage{graphicx}
\usepackage{txfonts}
\usepackage{longtable}
\usepackage[authoryear]{natbib}       
\bibpunct{(}{)}{;}{a}{}{,}
\begin{document}
  \title{Extended CO emission in the field of the light echo of
  V838 Monocerotis\thanks{Based on observations carried out with the IRAM
  30-meter telescope. IRAM is supported by INSU/CNRS (France), MPG
  (Germany), and IGN (Spain).}}

  \author{T. Kami\'{n}ski} 

  \offprints{T. Kami\'{n}ski}

  \institute{\center Department for Astrophysics, N. Copernicus
            Astronomical Center, 		  Rabia\'{n}ska 8,
            87-100 Toru\'{n}, Poland\\ 
            \email{tomkam@ncac.torun.pl}}
	    
  \date{Received; accepted}

\abstract
 {V838 Mon erupted at the beginning of 2002 becoming an extremely
 luminous star with L=10$^6$~L$_{\sun}$. The outburst was followed by
 the most spectacular light echo that
 revealed that the star is immersed in a diffuse and dusty
 medium, plausibly interstellar in nature. Low angular resolution
 observations of the star and its closest vicinity in the lowest
 CO rotational transitions revealed a molecular emission from the
 direction of V838~Mon. The origin of this CO emission has not been
 established.}  
 {The main aim of this paper is to better constrain the nature of the
 CO emission. In particular, we investigate the idea that the
 molecular emission originates in the material responsible for the
 optical light echo.}  
 {We performed observations of 13 positions within the light echo in
 the two lowest rotational transitions of $^{12}$CO using the IRAM
 30~m telescope.} 
 {Emission in CO $J=\,$1--$\,$0 and $J=\,$2--1 was detected in three
 positions. In three other positions only weak $J=\,$1$-\,$0 lines
 were found. The lines appear at two different velocities $V_{\rm
 LSR}=\,$53.3~km~s$^{-1}$ and $V_{\rm LSR}=\,$48.5~km~s$^{-1}$, and both
 components are very narrow with {\small FWHM}$\,\approx\,$1~km~s$^{-1}$.}
 {The molecular emission from the direction of V838~Mon is extended and has a
 very complex distribution. We identify the emission as arising from
 diffuse interstellar clouds. A rough
 estimate of the mass of the molecular matter in those regions gives a
 few tens of solar masses. The radial velocity of the emission
 at 53.3~km~s$^{-1}$ suggest 
 that the CO-bearing gas and the echoing dust are collocated in the
 same interstellar cloud.} 


\titlerunning{CO emission in the field of the light echo}

\authorrunning{T. Kami\'{n}ski}
 
\maketitle

\section{Introduction \label{intro}}

V838~Mon is a star that brightened significantly at the
beginning of 2002 becoming a very luminous object
with a luminosity of 10$^6$~L$_{\sun}$. The outburst lasted
for $\sim$3~months and was characterized by a rather complex light curve
with three distinct maxima seen in the optical. Several broad P-Cygni profiles
observed during the brightening imply that the event was accompanied
by an outflow with terminal velocities of several
hundred~km~s$^{-1}$ \citep[e.g.][]{kip04}. The end of the eruption was
characterized by a 
very steep decline with a total drop in brightness in 
the V band of about 8~mag within one month. Unlike novae, the object
evolved from 
a hot stage with an F-type spectrum to progressively lower and lower
temperatures and eventually it turned into a supergiant cooler
than M10 \citep{eva03}. Named the first L supergiant, the
star has exhibited an extraordinary molecular 
spectrum studied extensively in optical \cite[e.g.][]{pav06} and
infrared \citep[e.g.][]{lyn04}. Even five years after 
the eruption, the object continues to evolve - after four years
of a relatively constant photometric brightness, an eclipse-like event
has been observed in the B and V light curves \citep{mun07}. Detailed
descriptions of the spectral and photometric behavior of V838~Mon can
be found in: \cite{mun02}, \cite{kim02}, \cite{wis03},
\cite{cra03,cra05}, \cite{kip04}, and \cite{tyl05}.  

Although there is great interest in V838~Mon among astrophysicists
and many efforts have been made to gain a
deeper understanding of the object, its nature remains unclear. A wide
range of mechanisms have been proposed to explain the outburst
including {\it (i)} thermonuclear models:  
a nova-like \citep{mun02} or a He-shell flash \citep{law05, munhan05}, and {\it
  (ii)} merging events: a stellar merger of stars with masses of
8~M$_{\sun}$ and 0.3~M$_{\sun}$ \citep{soktyl03,soktylconf}, and a
giant swallowing planets \citep{ret03,ret06}. It seems that the most promising
is the stellar merger scenario \citep{tylsok06,soktylconf}.     
  
Since February 2002 a light echo of V838~Mon has been observed
\citep{han02}. Its spectacular evolution, especially well documented
by the series of the Hubble Space Telescope (HST) images
\citep{bond03,bondconf}, revealed that V838~Mon is immersed in a
diffuse and dusty medium with a very complex spatial
distribution. There is no agreement  
about the origin of this matter. An analysis of the light echo images
performed by \cite{tyl04} shows that the dusty region has a rather complex distribution  and, most probably, is of interstellar origin \citep[see
  also][]{cra05,tyl05b}. On the other hand, there are suggestions that
the matter might have been ejected by the object itself in the past
\citep{bond03,bondconf}. 

A light echo analogue was discovered in the infrared with the Spitzer
    Space Telescope \citep{baner06}. The analysis of
    the infrared emission gives a few tens to a few hundreds of solar masses as
an order of magnitude estimate for the mass of the gas associated with
the emitting dust. This result is a strong argument supporting the idea
of an interstellar origin of the echoing matter.       

\cite{kam07a,kam07b} discovered emission in the $^{12}$CO
(2--1) and (3--2) transitions at the position of V838~Mon. The emission was
detected inside the large KOSMA-telescope beams with half
power widths (HPBWs) of 130$\arcsec$ and 82$\arcsec$. The detected
lines are very narrow and appear at LSR velocity of
53.3~km~s$^{-1}$. The data analysis
performed in \cite{kam07b} suggest that the emission is extended, but
on the basis of the low-resolution observations it was not
certain. Moreover, in observations performed  
in three epochs Kami\'{n}ski et al. found small but possibly real
variations in the intensity of the CO (2--1) line. Due to the unknown
contribution to the intensity changes from antenna-pointing errors, the
finding is also very uncertain. 

\cite{degu07} observed the region around V838~Mon in the CO (1--$\,$0)
transition and detected a narrow emission line 30$\arcsec$ north from the
star position. This detection together with the observations reported
in \cite{kam07b} show that some form of molecular
matter must exist close to the direction of V838~Mon. Its nature has
not been sufficiently established, but it is tempting to link the CO
emitting gas with the dusty medium illuminated by the eruption of
V838~Mon.

In the current paper, we present follow-up observations of the field around
V838~Mon in the CO (1--$\,$0) and (2--1) rotational transitions with a
much better angular resolution than the observations reported in
\cite{kam07a,kam07b}. The observational details are provided in
Sect.~\ref{obs}, while the data are described in Sect.~\ref{res}. In
Sect.~\ref{dis}, we discuss the results, in particular, we investigate
the possible origin of the CO emission found in the field around
V838~Mon. Sect.~\ref{sum} contains final conclusions and emphasizes
the need for future observations of the echo region in molecular
transitions.         

\section{Observations and data reduction \label{obs}}
\begin{figure}
 \resizebox{\hsize}{!}{\includegraphics[angle=270]{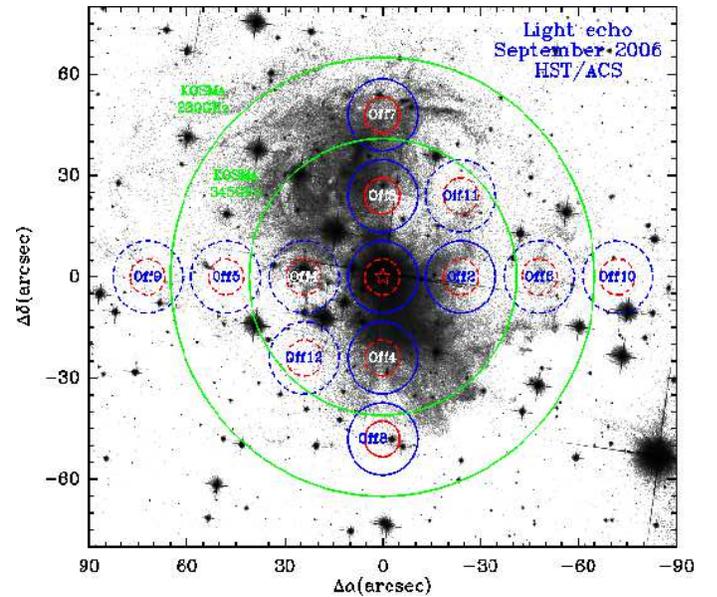}}  
 \caption{IRAM and KOSMA beams (HPBWs) overlayed on the HST/ACS
light echo image from 10 September 2006 obtained with the F814W filter (image
downloaded via MAST, {\tt \tiny
  http://archive.stsci.edu}). The two largest circles  
({\it green}) mark the KOSMA telescope beams
at 230~GHz and 345~GHz centered at the star position and they
represent the low angular resolution observations in CO carried out in
2005 and 2006. The 13 positions
observed with the IRAM 30~m telescope on 27-28 September 2006 are
represented by the set of smaller circles that correspond to the beams
at 115~GHz (21$\arcsec$, {\it blue}) and 230~GHz (11$\arcsec$, {\it
  red}). These points with their coordinates are listed  
in Table~\ref{tab1}. The beams drawn with a solid line mark those
positions where CO emission was detected, while the beams drawn with a
dashed line mark observed positions with no detection. The axis show
the offset scales with respect to the V838~Mon position, which is
marked with a star symbol.}  
\label{echo}
\end{figure}

We observed 13 positions within the light echo of V838~Mon,
including the star position. They are listed in
Table~\ref{tab1} and marked on the optical light echo image in
Fig.~\ref{echo}. Each of the positions was observed in
the two lowest rotational transitions of $^{12}$CO, i.e. $J\,=\,1$--$\,0$ at
2.6~mm (115.271~GHz) and $J\,=2\,$--1 at 1.3~mm (230.538~GHz).  
 
\begin{table}
\caption{Positions observed in CO $J=$1--$\,$0 and 2--1 in Sept. 2006.}
\label{tab1}
\begin{tabular}{lrrccc}
\hline
\hline
\multicolumn{1}{c}{position}&$\Delta\alpha$&$\Delta\delta$&int.
time&$\sigma_{\rm 
  rms}$(1$-\,$0)&$\sigma_{\rm rms}$(2--1)\\ 
&[$\arcsec$]&[$\arcsec$]&[min]&[mK]&[mK]\\[2pt]
\hline
&&&&&\\[-8pt]
V838 Mon&0&    0&353.6 &14.1&46.3 \\	 
Off1 &  24&    0& 66.3 &27.3&82.1 \\	 
Off2 &--24&    0& 55.2 &30.7&83.4 \\ 
Off3 &   0&   24& 55.2 &33.4&95.8 \\   
Off4 &   0& --24& 55.2 &30.4&88.4 \\   
Off5 &  48&    0& 55.2 &41.2&154.0\\   
Off6 &--48&    0& 44.2 &42.7&130.3\\   
Off7 &   0&   48& 55.3 &33.9&106.1\\   
Off8 &   0& --48& 55.2 &33.2&102.3\\   
Off9 &  72&    0& 22.1 &46.9&149.2\\   
Off10&--72&    0& 33.1 &36.2&117.3\\   
Off11&--24&   24& 33.2 &51.4&143.9\\   
Off12&  24& --24& 33.1 &50.4&165.7\\	 
\hline
\end{tabular}
\end{table}
\begin{figure}
\resizebox{\hsize}{!}{\includegraphics{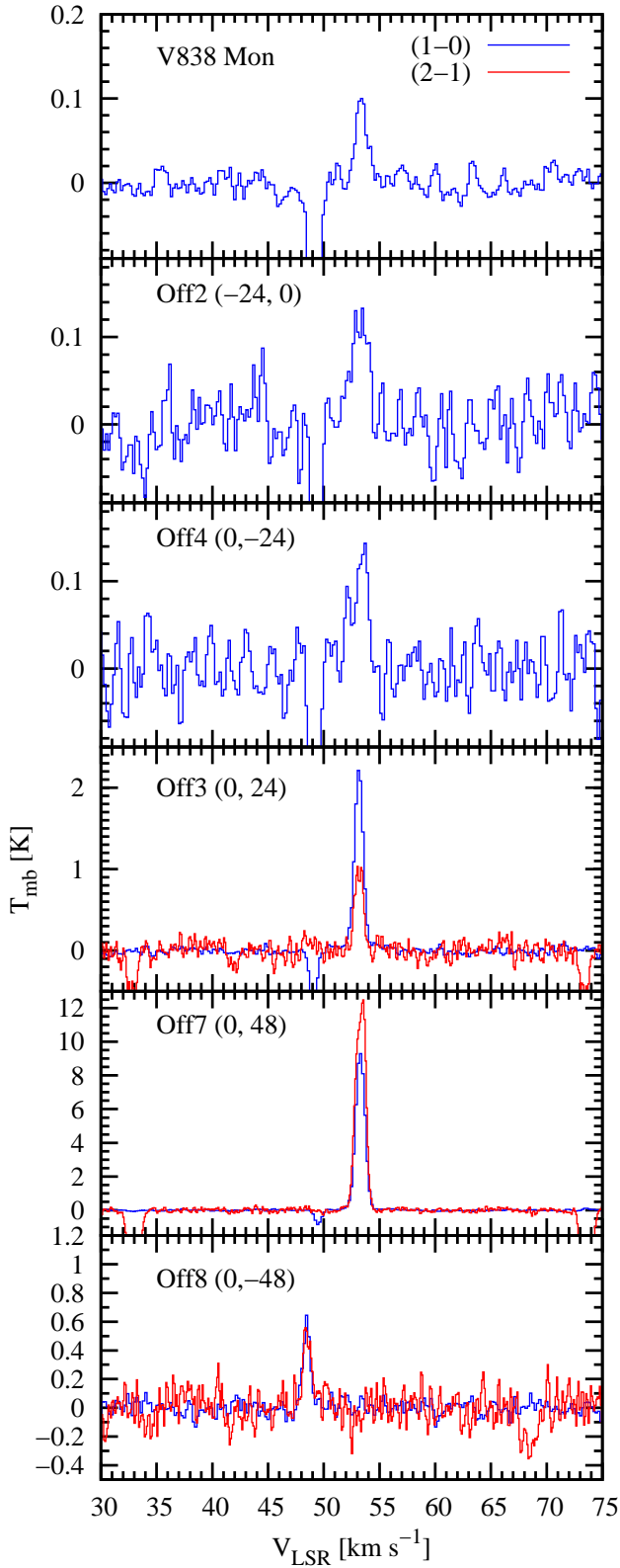}}  
\caption{Spectra with detected emission in $^{12}$CO $J=1$--0 ({\it
 blue}) and $J=2$--1 ({\it red}). The measured parameters of these
 emission lines are given in Table~\ref{tab2}. The strong 'absorption'
 feature present in the spectra of the $J=1$--0 transition at
 49~km~s$^{-1}$ is an alias of a mesospheric CO emission. In the
 $J=2$--1 spectra the 'absorption' features are aliases of the interstellar
 emission (they appear symmetrically on both sides of the
 emission). Since the alias of mesospheric emission affects the
 profile of interstellar CO (1--$\,$0) feature at the Off8 position in
 the folded spectrogram, we show the unfolded spectrum for this position.}  
\label{fig2}
\end{figure}

The observations were obtained with the IRAM 30~m telescope at Pico
Veleta, Spain, on 27-28 September 2006. The
half-power beam widths of the telescope are 
$21\farcs4$ and $10\,\farcs7$ at 115~GHz and 230~GHz, respectively. The
 spatial grid of the observations was arranged so that the distance
 between two neighboring positions is close to the twice the
 beam-width at 230~GHz, and the individual measurements obtained in
 the CO (2--1) transition can be considered as independent. 
 
 As frontends we used four (single pixel) heterodyne SIS receivers
(A100, B100, A230, and B230) simultaneously. The observations were
carried out in single sideband mode, with rejection factors of the
image sideband set to 22~dB and 
17~dB for the observations at 115~GHz and 230~GHz, respectively. As a
backend we used the spectrometric array VESPA. All the spectra were
acquired with a resolution of 78.1~kHz and with a bandwidth of
107~MHz. The frequency resolution corresponds to a resolution in
velocity units of 0.20~km~s$^{-1}$ at 115~GHz, and 0.10~km~s$^{-1}$ at
230~GHz.
 
All the observations were performed using a frequency switching
technique. The frequency switch throw was 15.6~MHz (i.e. frequency was
switched by $\pm$7.8~MHz), which is equivalent to 
a velocity throw of 40.7~km~s$^{-1}$ at 115~GHz, and 20.3~km~s$^{-1}$
at 230~GHz. Pointing and focusing were carried out on the strong continuum 
source 0528+134 every $\sim$2$\,$h. The resulting pointing
accuracy of our observations is better than 4$\arcsec$.

The data were calibrated using a chopper wheel method
\citep{kutner81} giving spectra in the antenna
temperature, $T_{\rm A}^\ast$, i.e. corrected for atmospheric attenuation,
ohmic losses, rearward spillover and scattering. The calibration at
the IRAM 30~m is known to be very accurate and should be 
better than 10\% for observations at 115~GHz.

The wavy baselines typical for frequency switching
observations were reduced by subtracting sinusoidal functions; in a
few cases subtraction of high order polynomials was necessary. All bad
channels were removed, spectra from two receivers with orthogonal
polarizations were coadded, and the resulting spectra were folded using the
standard shift-and-add method. Finally, the observations were
converted to the main beam temperature, $T_{\rm mb}$, using forward
and main beam efficiencies ($F_{\rm eff}$ 
and $B_{\rm eff}$ respectively) of 0.95 and 0.74 at 115~GHz, and of
0.91 and 0.52 at 230~GHz. All the data reduction and analysis was
performed using the {\small GILDAS}\footnote{See {\tt
    http://www.iram.fr/IRAMFR/GILDAS}} software package. 

Although in this paper the $T_{\rm mb}$ scale is used, one can easily
convert this intensity scale to ordinary flux density units using the
conversion factor $S_{\nu} / T_{\rm mb}$ of 4.91~Jy~K$^{-1}$ for
both the data sets obtained at 115~GHz and 230~GHz. All the velocities in this
paper are given with respect to the local standard of rest (LSR).    

Total integration times (a sum of the integration times from two
orthogonally polarized receivers) and noise levels of the folded
spectra are given in Table~\ref{tab1}. The noise levels are given as a
standard deviation ($\sigma_{\rm rms}$, in the main beam temperature)
of a linear fit to the folded spectrum excluding all emission features
and their aliases.      
 
\section{Results \label{res}}
Emission in the CO (1--$\,$0) transition was detected at 6 positions:
Off2, Off3, Off4, Off7, Off8, and
at the V838~Mon position; at three of those positions, namely at Off3,
Off7, and Off8, also CO (2--1) emission is clearly present. The
beams with a positive detection are shown with a solid line on
Fig.~\ref{echo}. The spectra with the detections are shown on
Fig.~\ref{fig2} and all 
the emission lines are characterized in Table~\ref{tab2} in terms of
their central velocities, full widths at half maximum, intensities of
the peak, and integrated intensities. The values given in
Table~\ref{tab2} are results of a single Gaussians fit to the line
profiles. All the reduced spectra are displayed in Figs.~\ref{apfig1}
and~\ref{apfig2} in the online Appendix~\ref{ap1}.  
\begin{table*}
\begin{minipage}[t]{\hsize}
\caption{Parameters obtained from single Gaussians fitted to the spectra.}
\label{tab2}
\centering \renewcommand{\footnoterule}{}  
\begin{tabular}{crr|cccc|cccc}
\hline
\hline
&&&&&&&&\\[-8pt]
\multicolumn{3}{c}{}&\multicolumn{4}{|c}{$^{12}$CO
  (1$-\,$0)}&\multicolumn{4}{|c}{$^{12}$CO (2--1)}\\ 
position&$\Delta\alpha$&$\Delta\delta$&V$_{\rm LSR}$&FWHM&$T_{\rm mb}$
peak&$I_{\rm CO}$&V$_{\rm LSR}$&FWHM&$T_{\rm mb}$ peak&$I_{\rm CO}$\\ 
&[$\arcsec$]&[$\arcsec$]&[km~s$^{-1}$]&[km~s$^{-1}$]&[K]&[K~km~s$^{-1}$]
&[km~s$^{-1}$]&[km~s$^{-1}$]&[K]&[K~km~s$^{-1}$]\\[2pt]    
\hline 
&&&&&&&&\\[-8pt]
V838Mon& 0   & 0   &53.4& 1.1&  0.10&  0.13&  &	    &	      &\\
Off2   & --24& 0   &53.2& 1.8&  0.12&  0.24&  &	    &	      &\\  
Off3   & 0   & 24  &53.2& 0.9&  2.22&  2.17& 53.2&0.9&1.03&0.97\\
Off4   & 0   & --24&53.3& 1.7&  0.12&  0.22&  &	    &	      &\\
Off7   & 0   & 48  &53.3& 1.0&  9.47&  9.77&53.3\footnote{the profile is clearly not a single Gaussian}&1.1$^a$&12.70$^a$&14.23$^a$\\ 
Off8   & 0   & --48&48.5\footnote{measurements obtained at the
  unfolded spectrum}&0.8$^b$&0.61$^b$&0.54$^b$&48.5&0.8&0.50&0.41\\ 
\hline
\end{tabular}
\end{minipage}
\end{table*}

The spectra displayed in Fig.~\ref{fig2} require a technical
comment. Since frequency switching was employed as the observing
method, the telluric (mesospheric) emission of CO was not removed from the
spectra \cite[see][for more detailes about telluric CO emission in
  frequency switching observations]{thum95}. The telluric emission
appears at the LSR velocity of the mesosphere in the time of the observations,
which in our case was typically $\sim$8.8~km~s$^{-1}$. After folding
procedure the telluric CO affects the spectra by the emission and its
two aliases, i.e. absorption-like features shifted by $\pm$7.8~MHz
with respect to the emission feature. Unfortunately, at the Off8
$(0,-48)$ position celestial emission appears at (unexpected) velocity of
48.45~km~s$^{-1}$, which closely coincides 
with the position of the absorption alias of the CO (1--$\,$0) telluric
emission at 49.34~km~s$^{-1}$. Luckily, the emission line at
48.45~km~s$^{-1}$ is not affected by the mesosphecic alias on the
spectrum before the folding procedure. Therefore on
Fig.~\ref{fig2} we present the spectrum before folding and all the
measurements given in Table~\ref{tab2} for the CO (1--$\,$0) line at
48.45 km~s$^{-1}$ were performed on the unfolded spectrum, which has however a 
poorer noise characteristics than the unfolded one by a factor
of~$\sqrt{2}$.

In the following, we consider the detected CO (1--$\,$0) emission in
two groups, namely weak ($T_{\rm mb}<$ 0.15~K at the peak) and
strong features (with peak much greater than 0.15~K). Weak emission is
found at Off2, Off4, and at the star position. The emission at all of
these positions appears at a velocity of
$\sim$53.3~km~s$^{-1}$ and does not have any detectable counterpart in the
corresponding CO (2--1) spectra. Emission in CO (2--1) is not seen,
most probably due to higher noise in the 230~GHz data. 

The strong emission of CO (1--$\,$0) is found at the three
positions along the north-south direction: Off3, Off7,
and Off8. Especially prominent is the emission at Off3 and
Off7. The lines both appear at radial velocity of
$\sim$53.2~km~s$^{-1}$. The emission found at the Off8 position, as 
has already been noted, appears at the unexpected radial velocity
48.45~km~s$^{-1}$. All the three strong CO (1--$\,$0) features
have counterparts in the (2--1) spectra. Ratios of those lines
are discussed in Sect.~\ref{prof}.  

As can be seen in Table~\ref{tab2}, all the detected lines are very
narrow. The weakest CO (1--$\,$0) features appear somewhat broader
than the strong lines. Note that the feature found at the star
position, where the integration time was much longer than at Off2 and
Off4, is only slightly broader that the strong lines, hence one can state
 that the broadening of the weak features is an effect of a lower
 signal to noise ratio of the spectra. In general, the profiles of the
 detected lines are very well described by a
Gaussian shape, except the most prominent line in our data, i.e. the CO
(2--1) feature at Off7, which has a slightly asymmetric peak. 

\section{Discussion \label{dis}}

The IRAM observations reported here confirm the suggestion made in
\cite{kam07b} that the CO emission discovered in the direction of V838~Mon 
in the low angular resolution KOSMA data, is extended and not
directly related to the object that erupted in 2002. Indeed, the
measurements indicate that the CO
emission has a very complex spatial distribution and is not only
limited to the position of V838~Mon. In the following
discussion we put some constraints on the origin of the molecular emission.
In particular, we investigate the possibility of a physical connection
of the CO emitting matter with the dusty environment revealed by the
light echo.      

\subsection{Origin of the emission: circumstellar vs. interstellar \label{ori}}

The strongest argument against circumstellar origin of the CO-bearing
gas is the narrowness of the emission lines. If the extended molecular
region originates as a result of a stellar mass loss before 2002, it
would produce emission features certainly broader and more complex
than the ones we see  in the two CO transitions
\citep[e.g.][]{coex}. Similarly, the emission found at the star 
position cannot be identified with the matter lost during and after
the 2002 eruption. Indeed, 
 numerous P-Cygni profiles observed during the outburst revealed
 matter expelled with velocities of several hundred~km~s$^{-1}$
\citep[e.g.][]{kip04} and the continuous outflow observed in V838~Mon
since the outburst has a terminal velocity of about 150~km~s$^{-1}$
\citep{munconf}. If any of this lost
matter radiates in the CO rotational transitions, the emission
should appear as a very broad feature with FWHM$\ga\,$150~km~s$^{-1}$.  

The narrowness of the detected CO lines indicates
that the molecular emission arises in an interstellar
medium. According to the classification in \cite{dis93} and on the
basis of line-width measurements 
(Table~\ref{tab2}) we can classify the molecular regions we see in the
CO emission as diffuse clouds. In light of the finding of 
\cite{afsar07}, namely that V838~Mon is a member of a B-stars
association, the narrow CO emission can be attributed to an
interstellar medium within the cluster. It may be the matter remaining
after the dissipation of most of the parent molecular material from
which the cluster had formed, much like the matter seen in the Pleiades
cluster. We cannot rule out, however, the possibility that we see foreground
and/or background clouds with respect to the location of the
cluster or the star itself. The radial location of the clouds is discussed
in more detail in Sect.~\ref{comp}. 

Using the well known X$_{\rm CO}$-factor method we can estimate the mass of
molecular matter inside the cloud. As found in \cite{liszt07}, the method
gives satisfactory results even for very diffuse clouds, i.e. those with very
small column densities of CO. The mass of molecular hydrogen can be
expressed as: 
\begin{equation}\label{eq1}
M_H=X_{\rm CO}\Omega d^2m_{H_2}I_{\rm CO},
\end{equation}
where X$_{\rm CO}$ is the conversion factor for the column density of
H$_2$ to integrated intensity of the CO (1--$\,$0) emission, $\Omega$ is the
solid angle of the emitting region, $d$ is the distance to the cloud,
and $I_{\rm CO}=\int{T_{\rm mb}dV}$ is the main beam integrated
intensity. Here we take $X_{\rm CO}$=2.8 \citep[in units of
  10$^{20}\,$cm$^{-2}$~K$^{-1}$~km$^{-1}$~s,][]{blo} although values
as high as 6 can be found in literature for clouds in the outer
Galaxy (e.g. \cite{kam07b} found $X_{\rm CO}$=5.4 for the dark clouds
identified in the vicinity of V838 Mon). We can rewrite Eq.~(\ref{eq1}) as: 
\begin{equation}\label{eq2}
M_H=0.4 \Omega d^2 I_{\rm CO} \; {\rm M}_{\sun},
\end{equation}  
where $\Omega$ is expressed in arcmin$^2$, $d$ in kpc, and $I_{\rm CO}$ in
K~km~s$^{-1}$. The solid angle of a single beam at 115~GHz is approximately
0.144~arcmin$^2$. We are interested in the total mass of
the molecular matter radiating in CO (1--$\,$0) in the six
positions. With the sum of  
all the integrated intensities of $\sim$13.1~K~km~s$^{-1}$ (see
Table~\ref{tab2}) and the distance to the star of 6.1~kpc
\citep{spar07} we get a total mass of about 28~${\rm
  M}_{\sun}$. The CO emission can be more extended than it appears in
our spatially limited measurements, so the total mass of
molecular matter is probably even higher. Moreover, the above value does
not account for a mass of atomic
gas, which can contribute considerably to the total mass of a diffuse
cloud. This rough estimate shows, however, that the mass of the
sampled regions is already quite high and cannot be interpreted as
matter expelled by the star. 

\subsection{Spatial complexity of the molecular gas \label{prof}}

The CO (2--1) emission
found in our data exhibit various intensities with
respect to the strength of the corresponding CO (1--$\,$0)
emission detected at the same position. The ratios of the integrated
intensities of the CO (1$-\,$0) to (2--1) lines, ${\mathcal
  R}_{10:21}$, are 2.22, 0.69, and 
1.31 for the positions Off3, Off7, and Off8, respectively. For a
homogenous distribution of interstellar gas one would expect the
CO (1$-\,$0) line to be stronger, since it was observed with the larger
beam. The wide variety in the values of 
${\mathcal R}_{10:21}$ in our data can be interpreted as a
result of a highly complex distribution of the emitting 
gas. In other words, that can be an effect of different beam filling
factors for observations at 115~GHz and 230~GHz. Sharp inhomogeneities
must then occur on the plane of the sky at angular scales at least
comparable to our beam-size at 230~GHz, i.e. at ranges of order
10$\arcsec$. This corresponds to spatial scales of $\sim$0.3~pc at a
distance of 6~kpc. Such a small scale structure is commonly observed 
in CO maps of diffuse molecular clouds (see
Sect.~\ref{ori}). Alternatively, the variety of ${\mathcal R}_{10:21}$
values can be explained by somewhat unusual population of CO levels in
the molecular medium, but this seems to be unlikely.    

\subsection{The two velocity components \label{comp}}

The kinematical characteristics of the CO emission (see
Table~\ref{tab2}) indicates that the 
strong emission found in Off3 and Off7, together with the weak emission
appearing in Off2, Off4, and in the star position, are all physically
related. The emission features found at these positions appear at
nearly the same velocity of 53.3$\pm$0.1~km~s$^{-1}$ 
(1$\sigma$). Moreover, the emission sampled in the CO
(1--$\,$0) data, is spatially continuous and forms one region elongated in
the north-south direction. Thus, the emission lines can be interpreted as
emerging from  the same diffuse cloud. The central velocity of this
molecular region is very close to the velocity of the SiO maser
emission observed from the direction of V838~Mon 
\citep[54~km~s$^{-1}$, ][]{degu05}, which is often considered as
the radial velocity of the star itself. If so, the emitting molecular
matter seen in our data should reside very close to the star (see also
Sect.~\ref{conecho}).  

The emission detected in the Off8 position is the only one centered at
48.5~km~s$^{-1}$. None of the spectra contains an emission component at 
intermediate velocities between 48.5~km~s$^{-1}$ and 53~km~s$^{-1}$,
hence there is no transition zone between those two kinematic regions
in the area of our measurements. This further 
indicates that the portions of gas emitting at the two distinct
velocities are not 
physically connected and they form separate molecular
complexes. Assuming standard rotation law of the Galaxy
 in the direction of V838~Mon, the gas at lower radial velocity 
should reside $\sim$1~kpc closer to the observer. The
emission that appears at 48.5~km~s$^{-1}$ may be considered as
emerging from a foreground molecular cloud with respect to the SiO
maser, if one trusts in the kinematical distances.

\subsection{Association with the light echo material? \label{conecho}}

The CO emission is located in the field of the light echo. The
question that naturally arises is whether the molecular emission is
physically connected with the light echo material. The issue is
worthwhile to consider, especially in the context of discussion about
the origin of the echoing dust (see Sect.~\ref{intro}). 

To answer to the above question the relative location of the dust and the
radiating molecules along the line of sight must be known. The
distance to the echo material has been established by 
a geometric analysis of the echo evolution. As can be found in
\citet[][see Fig.~3 
  therein]{bondconf}, the reflecting regions seen in the echo should be located within several pc from the star, so
their heliocentric distance should be about 6~kpc \citep{spar07}. The
distance to the CO-bearing gas can be only poorly constrained. Kinematical
distance to the component at 53~km~s$^{-1}$, well correlated spatially with the
echo on the plane of the sky, is $\sim$7~kpc, but due to streaming
motions the real distance can be $\sim$1~kpc higher or lower. Thus, the
association of the molecular gas with the dust cannot be ruled out.

The current understanding of V838~Mon seems to favor scenarios where
the star is considered as a very young object. If so, it should
exhibit systemic velocity very close to the velocity of the local
interstellar medium. If the star has the same velocity as the SiO
maser, then the local interstellar medium, including the light echo
material, would have a velocity of 54~km~s$^{-1}$. As already noted in
the preceding section, this value agrees very well with the velocity
of most of the CO emission. It suggests that the dust and molecular
gas are located in the same cloud. However, this suggestion should be treaded
cautiously, since, in the case of V838~Mon, the maser can have
different nature than in other SiO stellar sources \citep{degu07}, and
consequently its radial velocity can be different than the real
velocity of the star. 

Furthermore, in the case of common origin of the dust and CO-bearing
gas, one might expect that there is some overall correlation between the
dust and CO distribution. Our data have still too low angular
resolution and too poor coverage of the field to look for any
meaningful correlations in the both distributions, but some
general remarks can be made. As can be seen in Fig.~\ref{echo}, the CO
emitting region extends basically along the north-south direction,
similarly to the most prominent reflections seen in the light echo in
the epoch close to the radio observations. The optical echo is however
more extended in the east-west direction. One should remember that
the echo on a single epoch image shows only a thin part of the whole
dusty environment \citep[e.g.][]{tyl04}, while the CO emission
probes the total column 
density of the cloud along the line of sight. More appropriate would
be, though, to compare the map of molecular emission with the optical
pictures summed over the time of the observable evolution of the echo.  

In summary, the current data do not allow us to conclusively verify
whether the CO emission is associated with the light echo material,
but they make such a possibility very probable. To
verify the idea further, a direct measurement of the radial
velocity of the echoing gas or more reliable constraints on the
systemic velocity of V838~Mon are needed. A CO map with a good
sampling and covering a region larger than the size of the light echo
would be very helpful as well. 

\section{Concluding remarks \label{sum}}

We present observations towards 13 positions in the field of the light
echo of V838 Mon in the CO (1--$\,$0) and (2--1) transitions. The
measurements reveal an extended molecular region around the star at two
distinct radial velocities.  The CO emitting region is elongated in
the north-south direction and exhibits a very complex distribution on
the plane of the sky. We identify the CO lines as emerging from
diffuse interstellar clouds. No molecular emission that can be
associated with the star itself was detected.  

The possible association of the molecular emission
with the light echo material has been investigated. Although the CO emission
appears in the field of the light echo, its detailed spatial
distribution correlates only weakly with the light echo image. On the
other hand, the velocity of the CO emission agrees very well with the
velocity of the SiO maser discovered from the direction of V838~Mon,
making the collocation of the dust and CO-bearing gas probable. 

To more deeply investigate the origin of the molecular emission in the
field of the echo, more extended map of the region in
different molecular transitions is needed. A fully sampled map of the
echoing region would be helpful to draw more conclusive statements about the
suggested connection with the light echo material. 

\begin{acknowledgements} 
 The author thanks R. Tylenda and the anonymous referee for their useful
 comments and suggestions towards improving the paper.  The author is
 also grateful to M. Pu{\l}ecka
 for her help in obtaining the observations at the IRAM 30 m
 telescope. The research reported in this paper was supported from a
 grant no. N203~004~32/0448 financed by the Polish Ministry of Science
 and Higher Education. 
\end{acknowledgements}
\bibliographystyle{aa}

\begin{thebibliography}{}


   \bibitem[Af{\c s}ar \& Bond(2007)]{afsar07} Af{\c s}ar, M., \&
   Bond, H.~E.\ 2007, \aj, 133, 387  

   \bibitem[Banerjee et al.(2006)]{baner06}
   Banerjee, D. P. K., Su, K. Y. L., Misselt, K. A., \& Ashok ,
   N. M. 2006, \apj, 644, L57

   \bibitem[Bloemen et al.(1986)]{blo} 
   Bloemen, J.~B.~G.~M., Strong, A. W., Mayer-Hasselwander, H. A., et
   al.\ 1986, \aap, 154, 25  

   \bibitem[Bond et al.(2003)]{bond03}
   Bond, H. E., Henden, A., Levay, Z. G., et al. 2003, \nat, 422, 405

   \bibitem[Bond(2007)]{bondconf}
   Bond, H. E. 2007, in The Nature of V838 Mon and Its Light
   Echo, eds. R. L. M. Corradi \& U. Munari, ASP Conf. Ser.,
   363, 130 

   \bibitem[Crause et al.(2003)]{cra03} Crause, L.~A., Lawson, W.~A.,
   Kilkenny, D., et al. 2003, \mnras, 341, 785 

   \bibitem[Crause et al.(2005)]{cra05} Crause, L.~A., Lawson, 
   W.~A., Menzies, J.~W., \& Marang, F.\ 2005, \mnras, 358, 1352 

    \bibitem[Deguchi et al.(2005)]{degu05}
   Deguchi, S., Matsunaga, N., \& Fukushi, H. 2005, PASJ, 57, 933

   \bibitem[Deguchi et al.(2007)]{degu07} 
   Deguchi, S., Matsunaga, 
   N., \& Fukushi, H.\ 2007, in The Nature of V838 Mon and its Light
   Echo, eds. R. L. M. Corradi \& U. Munari, ASP Conf. Ser., 363, 81 

   \bibitem[van Dishoek et al.(1993)]{dis93} van Dishoek, E.~F.,
   Blake, G.~A., Draine, B.~T., \& Lunine, J.~I.\ 1993, in Protostars and
   Planets III, 163  

   \bibitem[Evans et al.(2003)]{eva03}
   Evans, A., Geballe, T. R., Rushton, M. T., et al. 2003, \mnras, 343,
   1054

   \bibitem[Henden et al.(2002)]{han02}
   Henden, A., Munari, U., \& Schwartz, M. 2002, IAU Circ. 7859

   \bibitem[Kami\'{n}ski et al.(2007a)]{kam07a}
   Kami{\'n}ski, T., Miller, M., Szczerba, R., \& Tylenda, R.\ 2007,
   in The Nature of V838 Mon and its Light Echo, eds. R. L. M. Corradi \& U.        Munari, ASP Conf. Ser. 363, 103   
  
   \bibitem[Kami\'{n}ski et al.(2007b)]{kam07b}
   Kami\'{n}ski, T., Miller, M., Tylenda, R. 2007, \aap, 475, 569

   \bibitem[Kimeswenger et al.(2002)]{kim02} Kimeswenger, S., Lederle,
   C., Schmeja, S., \& Armsdorfer, B.\ 2002, \mnras, 336, L43
 
   \bibitem[Kipper et al.(2004)]{kip04} 
   Kipper, T., Klochkova, V. G., Annuk, K., et al.\ 2004, \aap, 416, 1107 

   \bibitem[Kutner \& Ulich(1981)]{kutner81}
   Kutner, M. L., Ulich, B. L. 1981, \apj, 250, 341

   \bibitem[Lawlor(2005)]{law05} 
   Lawlor, T.~M.\ 2005, \mnras, 361, 695

   \bibitem[Liszt(2007)]{liszt07} 
   Liszt, H.~S.\ 2007, \aap, 476, 291 
 
   \bibitem[Lynch et al.(2004)]{lyn04} Lynch, D.~K., Rudy, R.~J.,
   Russel, R.~W., et al.\ 2004, \apj, 607, 460  

   \bibitem[Munari et al.(2002)]{mun02} 
   Munari, U., Henden, A., Kiyota, S., et al.\ 2002, \aap, 389, L51  

   \bibitem[Munari \& Henden(2005)]{munhan05} 
   Munari, U., \& Henden, A.\ 2005, in Interacting Binaries:
   Accretion, Evolution, and Outcomes, AIP Conf. Proceedings, 797, 331 

   \bibitem[Munari et al.(2007a)]{mun07} 
   Munari, U., Corradi, R. L. M., Henden, A., et al. 2007, \aap, 474, 585

   \bibitem[Munari et al.(2007b)]{munconf} Munari, U., Navasardyan, H.,
   \& Villanova, S.\ 2007, in The Nature of V838 Mon and its Light
   Echo, eds. R. L. M. Corradi \& U. Munari, ASP Conf. Ser.,
   363, 13
 
   \bibitem[Pavlenko et al.(2006)]{pav06} 
   Pavlenko, Y.~V., van Loon, J. Th., Evans, A., et al.\ 2006, \aap, 460, 245  

   \bibitem[Retter \& Marom(2003)]{ret03} Retter, A., \& Marom, A.\
   2003, \mnras, 345, L25
 
   \bibitem[Retter et al.(2006)]{ret06} Retter, A., Zhang, B., Siess,
   L., \& Levinson, A.\ 2006, \mnras, 370, 1573  

   \bibitem[Soker \& Tylenda(2003)]{soktyl03} Soker, N., \& Tylenda,
   R.\ 2003, \apjl, 582, L105  
  
   \bibitem[Soker \& Tylenda(2007)]{soktylconf} Soker, N., \& Tylenda,
   R.\ 2007, in The Nature of V838 Mon and its Light Echo,
   eds. R. L. M. Corradi \& U. Munari, ASP Conf. Ser., 363,
   280
  
   \bibitem[Sparks et al.(2007)]{spar07} 
   Sparks, W.~B., Bond, H. E., Cracraft, M., et al.\ 2007, ArXiv
   e-prints, 711, arXiv:0711.1495     

   \bibitem[Teyssier et al.(2006)]{coex} 
   Teyssier, D., Hernandez, R., Bujarrabal, V., Yoshida, H., \&
   Phillips, T.~G.\ 2006, \aap, 450, 167 
   
   \bibitem[Thum et al.(1995)]{thum95} 
   Thum, C, Sievers, A., Navarro, S., Brunswig, W., Pe\~{n}alver, J.\ 1995,
   IRAM Tech. Report 228/95. 
  
   \bibitem[Tylenda(2004)]{tyl04}
   Tylenda, R.\ 2004, \aap, 414, 223

   \bibitem[Tylenda(2005)]{tyl05} Tylenda, R.\ 2005, \aap, 436, 1009 

   \bibitem[Tylenda \& Soker(2006)]{tylsok06} Tylenda, R., \& Soker,
   N.\ 2006, \aap, 451, 223 

   \bibitem[Tylenda et al.(2005)]{tyl05b} Tylenda, R., Soker, N., \&
   Szczerba, R.\ 2005, \aap, 441, 1099  

   \bibitem[Wisniewski et al.(2003)]{wis03} Wisniewski, J.~P.,
   Morrison, N.~D., Bjorkman, K.~S., et al.\ 2003, \apj, 588, 486 

\end{thebibliography}

\Online
\begin{appendix}\section{Presentation of all the spectra}\label{ap1}
  \begin{figure*}
  \centering
  \includegraphics[width=16.4cm,clip, trim = 20 50 30 40]{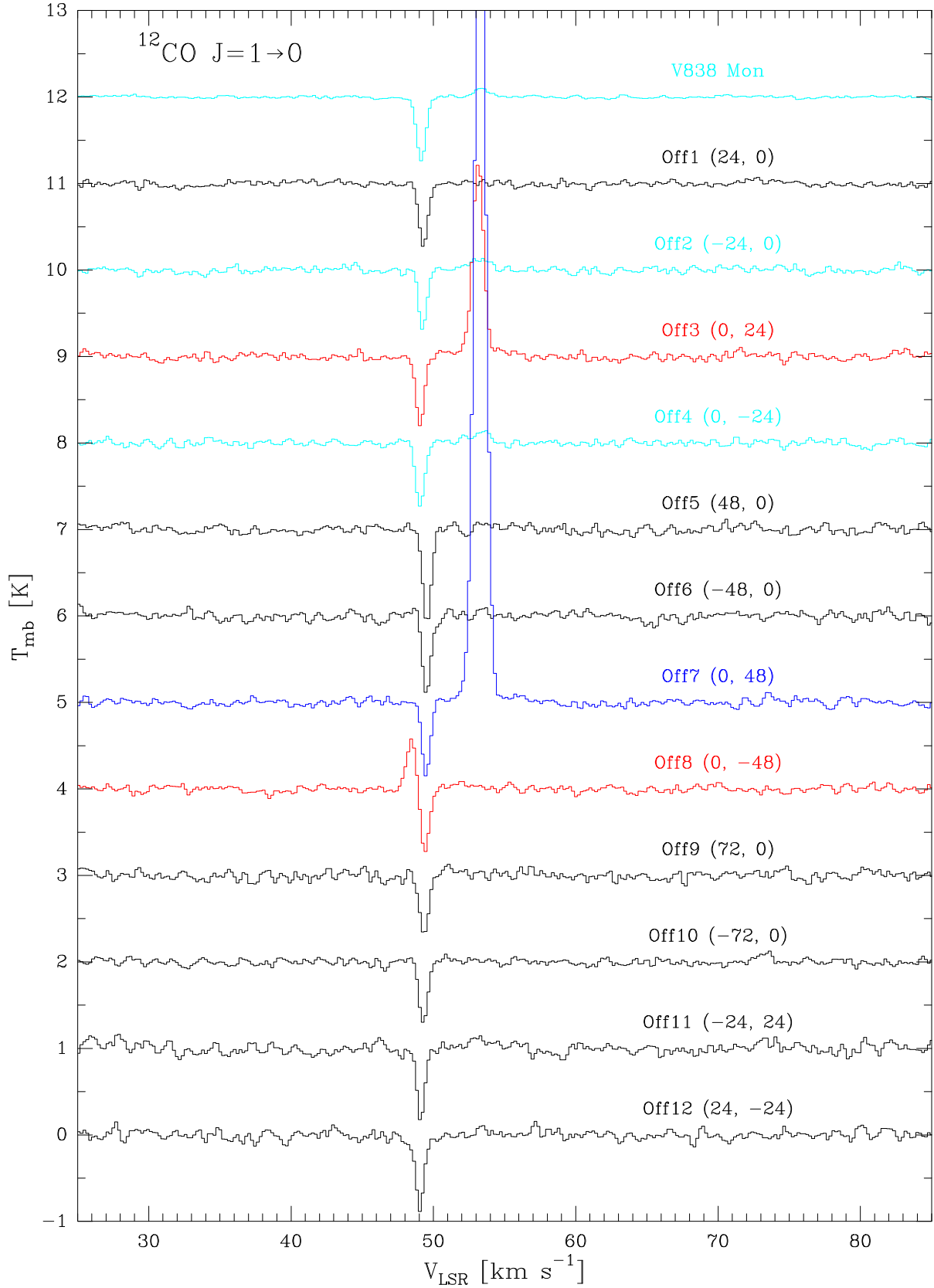}
  \caption{The folded spectra in $^{12}$CO $J\,=\,1-\,0$ for all
  the observed 13 positions. The 'absoprtion' feature at
  $\sim$49~km~s$^{-1}$ is an alias of the telluric CO
  emission. The spectra with positve detections are
  plotted with color lines.}\label{apfig1}
  \end{figure*}
  \begin{figure*}
  \centering
  \includegraphics[width=16.4cm,clip, trim = 20 50 30 40]{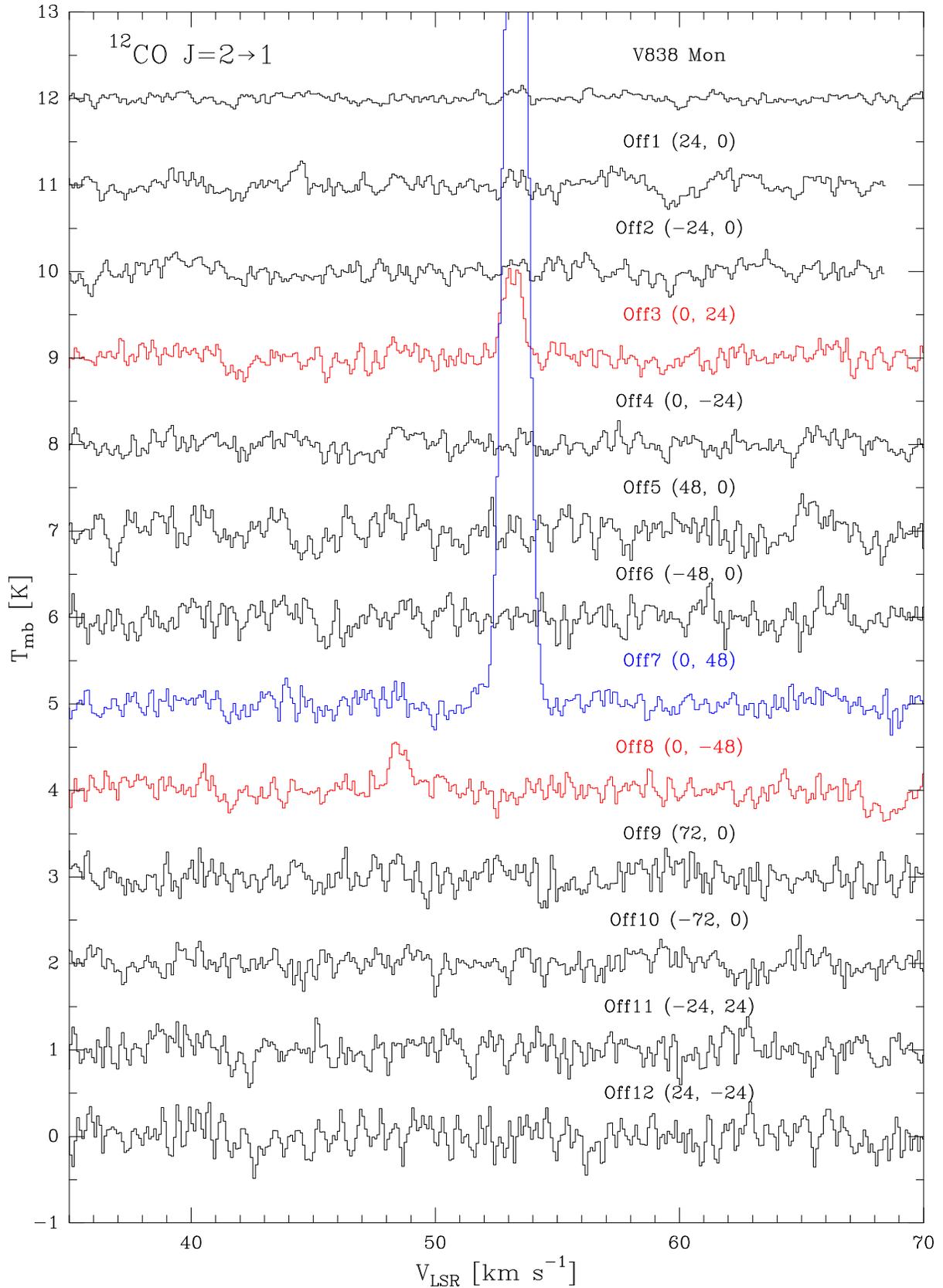}
  \caption{The folded spectra in $^{12}$CO
  $J\,=\,2-\,1$ for all the observed 13 positions. The spectra with
   positive detection are plotted with color lines.}\label{apfig2} 
  \end{figure*}
\end{appendix}
\end{document}